\documentclass{aa}
\usepackage{graphicx,hyperref,txfonts}

\begin{document}

\title{Model-independent characterisation of strong gravitational lenses}
\titlerunning{Model-independent characterisation of strong gravitational lenses}
\author{J. Wagner \and M. Bartelmann}
\institute{Universit\"at Heidelberg, Zentrum f\"ur Astronomie, Institut f\"ur Theoretische Astrophysik, Philosophenweg 12, 69120 Heidelberg, Germany\\ \email{j.wagner@uni-heidelberg.de}}
\date{Received XX; accepted XX}

\abstract{We develop a new approach to extracting model-independent information from observations of strong gravitational lenses. The approach is based on the generic properties of images near the fold and cusp catastrophes in caustics and critical curves. Observables used are the relative image positions, the magnification ratios and ellipticities of extended images, and time delays between images with temporally varying intensity. We show how these observables constrain derivatives and ratios of derivatives of the lensing potential near a critical curve. Based on these measured properties of the lensing potential, classes of parametric lens models can then easily be restricted to such parameter values compatible with the measurements, thus allowing fast scans of large varieties of models. Applying our approach to a representative galaxy (JVAS B1422+231) and a galaxy-cluster lens (MACS J1149.5+2223), we show which model-independent information can be extracted in those cases and demonstrate that the parameters obtained by our approach for known parametric lens models agree well with those found by detailed model fitting.}


\maketitle

\section{Introduction and motivation}
\label{sec:introduction}

Fitting a parametric gravitational-lens model to a given set of observed, gravitationally lensed images returns a set of parameter values that optimally reproduce the measured characteristics of the images with the given parametrised mass distribution. Such models are generally not unique because the same set of images can usually be fit by many different parametrisations. It is thus a question of conceptual and possibly practical importance as to what model-independent information is actually contained in strongly-lensed configurations of point-like or extended images. In fact, the only information we can infer on the deflector from the observables of strongly-lensed images is locally confined to the vicinity of these images. In this paper, we investigate which \emph{model-independent} information can be obtained from a given set of gravitationally lensed images. As we shall show, this information amounts to ratios of derivatives of the lensing potential on or near the critical curve.

In Sect.~\ref{sec:taylor}, we derive which model-independent information about the gravitational lens can generally be obtained from the mutual distances, the ellipticities and magnification ratios as well as time delays of multiply lensed images near fold and cusp points in critical curves. We further analyse the remaining degeneracies and estimate the measurement uncertainties and systematic errors of the results. In Sect.~\ref{sec:models}, we show how the parameters of parametrised mass models can be constrained by our approach. The allowed parameter ranges can then be compared to those obtained by direct model fitting. As representative example models, we consider axisymmetric and mildly elliptical models and investigate the influence of external shear on the ratios of derivatives. In Sect.~\ref{sec:examples}, we then extract the model-independent information from the bright triple images in the galaxy lens JVAS B1422+231 and the cluster lens MACS J1149.5+2223. Specialising our approach to lens models from the literature, we compare model parameters inferred from our approach with parameter values obtained by detailed model fitting. We summarise our results in Sect.~\ref{sec:summary}.

\section{Model-independent characterisation of gravitational lenses near folds and cusps}
\label{sec:taylor}

According to \cite{bib:Whitney}, the Fermat potential $\phi(x,y)$, $x, y \in \mathbb{R}^2$, of a sufficiently smooth gravitational lens model can be approximated around a singular point $(x^{(0)}, y^{(0)})$ by the fourth-order polynomial
\begin{align}
\phi_\mathrm{T}(x,y) &=
  \phi(0,0) + \tfrac12 y^2 - xy + \tfrac12 \phi_{11}^{(0)}x_1^2 + \tfrac16 \phi_{111}^{(0)}x_1^3 \label{eq:taylor_series} \\
  &+ \tfrac12 \phi_{112}^{(0)}x_1^2x_2 +  \tfrac12 \phi_{122}^{(0)}x_1x_2^2 +  \tfrac16 \phi_{222}^{(0)}x_2^3  \nonumber \\
  &+ \tfrac{1}{24} \phi_{1111}^{(0)}x_1^4  + \tfrac16 \phi_{1112}^{(0)}x_1^3x_2 +  \tfrac14 \phi_{1122}^{(0)}x_1^2x_2^2 \nonumber \\
  &+  \tfrac16 \phi_{1222}^{(0)}x_1x_2^3 + \tfrac{1}{24} \phi_{2222}^{(0)}x_2^4 \nonumber
\end{align}
if we introduce a coordinate system in the image plane with its origin shifted to $(x^{(0)},y^{(0)})$ and rotated such that
\begin{equation}
  \phi_{12}^{(0)} = 0 = \phi_{22}^{(0)}\;,
\label{eq:coordinate_system}
\end{equation}
and a coordinate system in the source plane such that
\begin{equation}
  \phi_1^{(0)} = 0 = \phi_2^{(0)}\;,
\label{eq:coordinate_system_2}
\end{equation}
without loss of generality. We further abbreviate
\begin{equation}
  \left.\frac{\partial\phi}{\partial x_i}\right\vert_{(x^{(0)},y^{(0)})} =
  \phi_i^{(0)}
\label{eq:abbrev}
\end{equation}
for $i=1,2$.

Given $\phi_\mathrm{T}(x,y)$, approximate lensing equations can be obtained from $\nabla_x\phi_\mathrm{T}(x,y) = 0$, where $\nabla_x$ denotes the gradient with respect to $x$. These approximate lensing equations are then simplified by keeping only the leading-order terms in $x$, as explained in \cite{bib:SEF} or \cite{bib:Petters}.

\subsection{Folds}
\label{sec:taylor_folds}

At a fold singularity, coordinate systems can be chosen such that the conditions
\begin{equation}
  \phi_1^{(0)} = \phi_2^{(0)} = \phi_{12}^{(0)} = \phi_{22}^{(0)} = 0
  \;,\quad
  \phi_{11}^{(0)} \ne 0 \;,\quad \phi_{222}^{(0)} > 0
\label{eq:conditions_fold}
\end{equation}
hold without loss of generality. In such coordinates, the approximate lensing equations to leading order in $x$ read
\begin{align}
  y_1 &= \phi_{11}^{(0)}x_{1} + \dfrac12\phi_{122}^{(0)}x_{2}^2 + \phi_{112}^{(0)}x_{1}x_{2}\;,
  \label{eq:fold_le1} \\
  y_2 &= \dfrac12\phi_{112}^{(0)}x_{1}^2 + \phi_{122}^{(0)}x_{1}x_{2} + \dfrac12\phi_{222}^{(0)}x_{2}^2\;.
  \label{eq:fold_le2}
\end{align}
Evaluating these equations for both images of a source at $y = (y_1, y_2)$, with image positions at $x_A$ and $x_B$, a system of lensing equations can be set up and solved for the derivatives of $\phi$ at $x^{(0)}$ by eliminating $y$.

If available, we can use the observed ratios of the semi-major to the semi-minor axis of the images,
\begin{equation}
  r_i = \dfrac{\phi_{22}^{(i)}}{\phi_{11}^{(0)}} \;,\quad i = A, B\;,
\end{equation}
to leading order in $x$ and the equation for the time delay between the two images to obtain
\begin{align}
  \phi_{222}^{(0)} &= \dfrac{12ct_\mathrm{d}^{(AB)}D_{\mathrm{ds}}}{D_\mathrm{d}D_\mathrm{s}  (1+z_\mathrm{d})(\delta_{AB2})^3}\;,
  \label{eq:time_delay_fold} \\
  \dfrac{\phi_{222}^{(0)} }{\phi_{11}^{(0)}} &= \dfrac{2r_A}{\delta_{AB2}} \label{eq:ratio_fold}
\end{align}
where $c$ denotes the speed of light, $z_\mathrm{d}$ the redshift of the lens plane, $t_\mathrm{d}^{(AB)}$ the measured time delay, $D_\mathrm{ds}$, $D_\mathrm{d}$, and $D_\mathrm{s}$ the angular diameter distances between the lens and the source planes, the observer and the lens, and the observer and the source, respectively. $\delta_{AB2} = x_{A2} - x_{B2}$ is the separation between (the centres of light of) the two images $A$ and $B$ at a fold in the lens plane. In the chosen coordinate system, the line connecting the two images is perpendicular to the critical curve, \cite{bib:SEF} (detailed derivations can be found in the Appendix).

The parity of the images can be determined by noting that the image leading in time has positive parity, while the following image has negative parity. Since the magnifications are equal for both images near a fold no information can be gained to leading order from the magnification ratio.

For a physical interpretation of the ratios of derivatives of the lens potential, we rewrite Eq.~\ref{eq:ratio_fold} in terms of convergence, shear and flexion
\begin{eqnarray}
\kappa_0 =& 1 - \dfrac12 \left( \phi_{11}^{(0)} + \phi_{22}^{(0)} \right)\;, \quad &\gamma_{1} = \dfrac12 \left( \phi_{22}^{(0)} - \phi_{11}^{(0)} \right)\;, \\
F_{1} =& \dfrac12 \left( \phi_{111}^{(0)} + \phi_{122}^{(0)} \right)\;, \quad &F_{2} = \dfrac12 \left( \phi_{112}^{(0)} + \phi_{222}^{(0)} \right)\;, \\
G_{1} =& \dfrac12 \left( \phi_{111}^{(0)} - 3\, \phi_{122}^{(0)} \right)\;, \quad &G_{2} = \dfrac12 \left( 3\, \phi_{112}^{(0)} - \phi_{222}^{(0)}  \right)
\end{eqnarray} 
to obtain in the chosen coordinates
\begin{equation}
  \dfrac{\phi_{222}^{(0)} }{\phi_{11}^{(0)}} = \dfrac{3F_2 - G_2}{4(1 - \kappa_0)}  = \dfrac{2r_A}{\delta_{AB2}}\;.
\end{equation}

\subsection{Cusps}
\label{sec:taylor_cusps}

At a cusp singularity, we introduce coordinates such that
\begin{equation}
  \phi_1^{(0)} = \phi_2^{(0)} = \phi_{12}^{(0)} = \phi_{22}^{(0)} = \phi_{222}^{(0)} = 0
\label{eq:conditions_cusp_1}
\end{equation}
as well as
\begin{equation}
  \phi_{11}^{(0)} \ne 0 \;,\quad \phi_{122}^{(0)} < 0 \;,\quad
  \phi_{2222}^{(0)} > 0
\label{eq:conditions_cusp}
\end{equation}
and
\begin{equation}
  (\phi_{122}^{(0)})^2 -\dfrac13\phi_{2222}^{(0)}\phi_{11}^{(0)} \ne 0
\label{eq:conditions_cusp}
\end{equation}
hold. Again, this is possible without loss of generality, \cite{bib:SEF}. We label the three images such that image $A$ is closest to the cusp inside the critical curve and has negative parity, while $B$ and $C$ have positive parity and fall above and below the critical curve, respectively. The image coordinates then satisfy
\begin{eqnarray}
 x_{A1} >& 0  \;, \quad &x_{A2} \ge 0  \;, \\  
x_{B1} \ge& 0  \;, \quad &x_{B2} > 0  \;, \\
x_{C1} \ge& 0  \;, \quad &x_{C2} < 0  \;.
\end{eqnarray}
The configuration with opposite parities can be calculated analogously. The observed image configuration is degenerate with respect to the parity of their images until time delay information is included to decide which of the images is leading in time and thus has positive parity. This implies that $\phi_{122}^{(0)}$ and $\phi_{2222}^{(0)}$ are only determined up to their signs without time delay information. Hence, we choose $\phi_{122}^{(0)} < 0$ and $\phi_{2222}^{(0)} > 0$ to fix the signs in the lensing equations.

Then, the Taylor-expanded lensing equations to leading order in $x$ read
\begin{align}
  y_1 &= \phi_{11}^{(0)}x_1 + \dfrac{1}{2} \phi_{122}^{(0)}x_2^2\;, \label{eq:cusp_le1} \\
  y_2 &= \phi_{122}^{(0)}x_1x_2 + \dfrac{1}{6}\phi_{2222}^{(0)}x_2^3\;.
\label{eq:cusp_le2}
\end{align}
Using the same notation for the constants and observables as for the folds, we find
\begin{align}
  \phi_{2222}^{(0)} &= \dfrac{8ct_\mathrm{d}^{(ij)}D_{\mathrm{ds}}}{D_\mathrm{d} D_\mathrm{s} (1 + z_\mathrm{d}) (\delta_{ij2})^4} \quad (i, j = A, B, C, \; i \ne j)\;,
\label{eq:time_delay_cusp} \\
  \dfrac{\phi_{122}^{(0)}}{\phi_{11}^{(0)}} &=  \dfrac{F_1 - G_1}{4(1 - \kappa_0)} = \dfrac{2\left( \delta_{AB1}\delta_{AC2} - \delta_{AC1}\delta_{AB2} \right)}{\delta_{AB2}\delta_{AC2}\left( \delta_{AB2} - \delta_{AC2} \right)}\;,
\label{eq:ratio1_cusp} \\
  \dfrac{\phi_{2222}^{(0)}}{\phi_{11}^{(0)}} &= \dfrac{2}{(\delta_{ij2})^2}\left( \dfrac{\phi_{122}^{(0)}}{\phi_{11}^{(0)}}\delta_{ij1} -r_i + r_j \right)\;;
\label{eq:ratio2_cusp}
\end{align}
these expressions are derived in the Appendix.

If the images are extended and time delays are available, Eqs.~\ref{eq:time_delay_cusp} and \ref{eq:ratio2_cusp} can be combined to determine $\phi_{11}^{(0)}$. As the coordinate differences $\delta_{ijk}$, $i, j = A, B, C$, $k = 1, 2$ between the images are not observable, we express them in terms of the measurable angles enclosed by the lines connecting $A$, $B$, and $C$,
\begin{align}
\delta_{AB1} =& -\delta_{AB} \, \cos\left(\dfrac{\alpha_A}{2}\right)\;, &\delta_{AB2} =& -\delta_{AB} \, \sin \left(\dfrac{\alpha_A}{2}\right) \;, \label{eq:cusp_trigonometry1}\\
\delta_{AC1} =& -\delta_{AC} \, \cos\left(\dfrac{\alpha_A}{2}\right)\;, &\delta_{AC2} =& \phantom{-} \delta_{AC} \, \sin \left(\dfrac{\alpha_A}{2}\right) \;, \\
\delta_{BC1} =& - \delta_{BC} \, \cos\left(\dfrac{\alpha_A + \alpha_B}{2}\right)\;, &\delta_{BC2} =& \delta_{BC} \, \sin\left(\alpha_B + \dfrac{\alpha_A}{2}\right) \label{eq:cusp_trigonometry2}
\end{align}
with $\alpha_i$, $i=A, B, C$, denoting the angles at the vertices $i$ of the image triangle.

Even if magnification ratios are prone to large uncertainties, we consider using them, as they allow to solve for the absolute position of one image. Inserting the image position into the lensing equations Eqs.~\ref{eq:cusp_le1} and \ref{eq:cusp_le2}, the source position can be determined. The latter, in turn, can be used to estimate the effect of truncating the Taylor approximation (as further detailed in Sect.~\ref{sec:taylor_error}) or to calculate the image positions assuming a certain lens model. This allows to test whether a given model describes an observed image configuration or to predict positions of further images not located in the vicinity of the critical curve.

Without loss of generality, we determine $x_{A}$, starting from the system of equations for the observable magnification ratios $\mu_{AB}$ and $\mu_{AC}$,
\begin{align}
  \mu_{AB} \equiv \dfrac{\mu_B}{\mu_A} &=
  \dfrac{r_{122}^{(0)}x_{A1} + \left(
      3r_{2222}^{(0)} - \left(r_{122}^{(0)}\right)^2
    \right)x_{A2}^2}{r_{122}^{(0)}x_{B1} + \left(
      3r_{2222}^{(0)}- \left(r_{122}^{(0)}\right)^2
    \right)x_{B2}^2} \;, \\
  \mu_{AC} \equiv \dfrac{\mu_C}{\mu_A} &=
  \dfrac{r_{122}^{(0)}x_{A1} + \left(
    3r_{2222}^{(0)} - \left(r_{122}^{(0)}\right)^2
  \right)x_{A2}^2}{r_{122}^{(0)}x_{C1} + \left(
    3r_{2222}^{(0)} - \left(r_{122}^{(0)}\right)^2
  \right)x_{C2}^2} \;,
\end{align}
where the ratios $r_{122}^{(0)}$ and $r_{2222}^{(0)}$ are given by the right-hand sides of Eqs.~\ref{eq:ratio1_cusp} and \ref{eq:ratio2_cusp}, respectively. Using the coordinate distances $\delta_{ijk}$ from Eqs.~\ref{eq:cusp_trigonometry1} to \ref{eq:cusp_trigonometry2} to replace $x_{Bi}$ and $x_{Ci}$, $i=1,2$, we can solve for $x_{A}$ and obtain
\begin{align}
  x_{A1} &= - \dfrac{\mu_{AB}\delta_{AB1}}{1 - \mu_{AB}} - \dfrac{3r_{2222}^{(0)} - \left(r_{122}^{(0)}\right)^2}{r_{122}^{(0)}} \nonumber \\ &\cdot
  \left(x_{A2}^2 + \dfrac{2\mu_{AB}\delta_{AB2}}{1 - \mu_{AB}}x_{A2} - \dfrac{\mu_{AB}\delta_{AB2}^2}{1-\mu_{AB}} \right) \label{eq:magnifications_cusp1} \;, \\
  x_{A2} &= \dfrac{-\left( \mu_{AC}\delta_{AC1} -  \tfrac{1-\mu_{AC}}{1-\mu_{AB}}\mu_{AB}\delta_{AB1} \right)}{2\left( \mu_{AC}\delta_{AC2} - \tfrac{1-\mu_{AC}}{1-\mu_{AB}} \mu_{AB}\delta_{AB2} \right)}\dfrac{r_{122}^{(0)}}{3r_{2222}^{(0)} -\left(r_{122}^{(0)}\right)^2} \nonumber \\ &-
  \dfrac{\tfrac{1-\mu_{AC}}{1-\mu_{AB}}\mu_{AB}\delta_{AB2}^2 - \mu_{AC}\delta_{AC2}^2}{2\left( \mu_{AC}\delta_{AC2} - \tfrac{1-\mu_{AC}}{1-\mu_{AB}} \mu_{AB}\delta_{AB2} \right)}\;.
\label{eq:magnifications_cusp2}
\end{align}
Table~\ref{tab:summary} summarises the model-independent information that can be determined for the different combinations of given observables. 

\begin{table}[t]
 \caption{Model-independent information that can be determined for different combinations of observables at folds and cusps: $\delta$ denotes the relative distances between the images (directly measurable in the fold case and determined by Eqs.~\ref{eq:cusp_trigonometry1} to \ref{eq:cusp_trigonometry2} at a cusp), $r$ the axis ratios of the extended images, $t_\mathrm{d}$ time-delay information, and $\mu$ magnification ratios.}
\label{tab:B1422}
\begin{center}
\begin{tabular}{lll}
\hline
\noalign{\smallskip}
 Observables & Fold & Cusp \\
\noalign{\smallskip}
\hline
\noalign{\smallskip}
 $\delta$ & -- & Eq.~\ref{eq:ratio1_cusp} \smallskip \\[0.2cm]
$\delta, t_\mathrm{d}$ & Eq.~\ref{eq:time_delay_fold}  & Eqs.~\ref{eq:time_delay_cusp}, \ref{eq:ratio1_cusp} \smallskip \\
$\delta, r$ & Eq.~\ref{eq:ratio_fold} & Eqs.~\ref{eq:ratio1_cusp}, \ref{eq:ratio2_cusp} \smallskip \\[0.2cm]
$\delta, t_\mathrm{d}, r $ & Eq.~\ref{eq:time_delay_fold}, \ref{eq:ratio_fold}  & Eqs.~\ref{eq:time_delay_cusp}, \ref{eq:ratio1_cusp}, \ref{eq:ratio2_cusp} \smallskip \\
$\delta, r, \mu$ & Eq.~\ref{eq:ratio_fold} & Eqs.~\ref{eq:ratio1_cusp}, \ref{eq:ratio2_cusp} , \ref{eq:magnifications_cusp1}, \ref{eq:magnifications_cusp2} \smallskip \\[0.2cm]
$\delta, t_\mathrm{d}, r, \mu$ & Eq.~\ref{eq:time_delay_fold}, \ref{eq:ratio_fold} & Eqs.~\ref{eq:cusp_le1}, \ref{eq:cusp_le2}, \ref{eq:ratio1_cusp}, \ref{eq:ratio2_cusp} , \ref{eq:magnifications_cusp1}, \ref{eq:magnifications_cusp2} \smallskip \\
\noalign{\smallskip}
\hline
\end{tabular}
\end{center}
\label{tab:summary}
\end{table}

\subsection{Uncertainties, errors and degeneracies}
\label{sec:taylor_error}

Each (ratio of) potential derivatives in Sect.~\ref{sec:taylor} is subject to measurement uncertainties, a possible systematic error from signal processing, and a systematic deviation from the true value due to truncating the Taylor approximation after the leading order. 

Statistical and systematic uncertainties can be propagated as usual, if given. Otherwise, calculating the results for all possible combinations of observables yields a range of values whose width indicates their uncertainties, because we expect the results to be independent of the specific image pair they are derived from. For example, by Eq.~\ref{eq:ratio_fold}, $\phi_{222}^{(0)}/\phi_{11}^{(0)}$ can be calculated from the axis ratios of both images $A$ and $B$. The difference between the two results is an estimate for the combined observational and methodical uncertainties. For potential ratios at a cusp, the number of possible ways to derive the same quantity is increased by the third image, thus improving the uncertainty estimate.

The possible bias due to truncating the Taylor series of the potential is expected to decrease the closer the images are to the critical curve and the closer the source is to the caustic. At a cusp, these distances can be calculated as described in Sect.~\ref{sec:taylor_cusps}, if the required observables are available.

Since the accuracy of the Taylor approximation is model dependent, a specific lensing potential needs to be assumed for estimating it. For elliptical models (elliptical potentials or surface-mass densities with singular isothermal density profiles) of moderate ellipticity $\lesssim0.2$, we obtain deviations of a few percent for sources closer to the caustic than $\sim5\,\%$ of the maximum extent of the caustic. In this case, results from time delays deviate by $\sim0.1\,\%$ and ratios of potential derivatives by up to $3.5\,\%$. The lower accuracy of the latter is due to the Taylor-expanded lensing equations having been further linearised, which is not necessary for the time-delay equation (details about the calculations can be found in the Appendix). Our estimates for the accuracy of results from time delays agree with similar estimates by \cite{bib:Congdon}.

Furthermore, the possible bias due to the restriction to leading-order terms is negligible for axisymmetric and elliptical models because their symmetry implies that most of the omitted terms vanish.

As already pointed out by \cite{bib:Gorenstein} and further developed by \cite{bib:Sluse}, several continuous transformations can be applied to the lens modelling equations leaving the observables invariant. In our case, we still have the freedom to scale all derivatives of $\phi$ by a factor $\lambda \in \mathbb{R}$. This would only change the source position which is not observable. The ratios of the derivatives remain invariant, and only the time delay can be used to break the degeneracy.

\section{Model selection}
\label{sec:models}

While our approach to extracting model-independent information on strong gravitational lenses from the observables is new to our knowledge, numerous ways to constrain parameters for lens models have been developed in the past, like \cite{bib:Bartelmann, bib:Gorenstein, bib:Grossman, bib:Hammer, bib:Jullo, bib:Keeton, bib:Limousin, bib:Narayan1, bib:Narayan2, bib:Oguri, bib:Suyu}. To connect our work to previous studies, we shall now relate the (ratios of) potential derivatives to specific lens models in order to constrain their parameters.

For any gravitational lens producing one image pair at a fold singularity only, we can determine a single model parameter by means of Eq.~\ref{eq:ratio_fold} and use Eq.~\ref{eq:time_delay_fold} to break the scaling degeneracy discussed in Sect.~\ref{sec:taylor_error}. At a cusp singularity with three neighbouring images, we have Eqs.~\ref{eq:ratio1_cusp} and \ref{eq:ratio2_cusp} to determine up to two model parameters and break the scaling degeneracy with Eq.~\ref{eq:time_delay_cusp}.

If the number of parameters exceeds the number of equations, the system of equations is underdetermined and a family of model parameters satisfying the observational constraints is obtained as a solution set, unless the system is inconsistent due to contradictory observations, or further information about the lens is available from non-lensing measurements, e.g.\ from observed velocity dispersions along the line-of-sight. Multiple sets of images from different sources at several singular points allow to further narrow the range of feasible model parameters.

\subsection{Axisymmetric lens models}
\label{sec:axisymmetric_models}

As cusps in axisymmetric models always degenerate to a point singularity in the source plane, next to which sources form two images on opposite sides of the lens, the only applicable axisymmetric case for our approach are double images at radial critical curves. Hence, models with tangential critical curves only, such as the point mass or the singular isothermal sphere, can be excluded from the analysis. Furthermore, lying much closer to the lens centre than the tangential critical curves, images near radial critical curves are hard to detect and so far, only a few of them have been found; see \cite{bib:Molikawa} and \cite{bib:Meneghetti} for an overview of the current observational status. Despite the restricted number of viable axisymmetric models -- such as the non-singular isothermal sphere, the \cite{bib:Plummer}, Navarro-Frenk-White \citep{bib:NFW}, and \cite{bib:Hernquist} models -- and the small number of confirmed, observed radial arcs, this class of models may still prove useful for primary lens models when adding external shear, as detailed in Sect.~\ref{sec:external_shear}.

\begin{figure}[!ht]
\centering
  \includegraphics[width=0.49\textwidth]{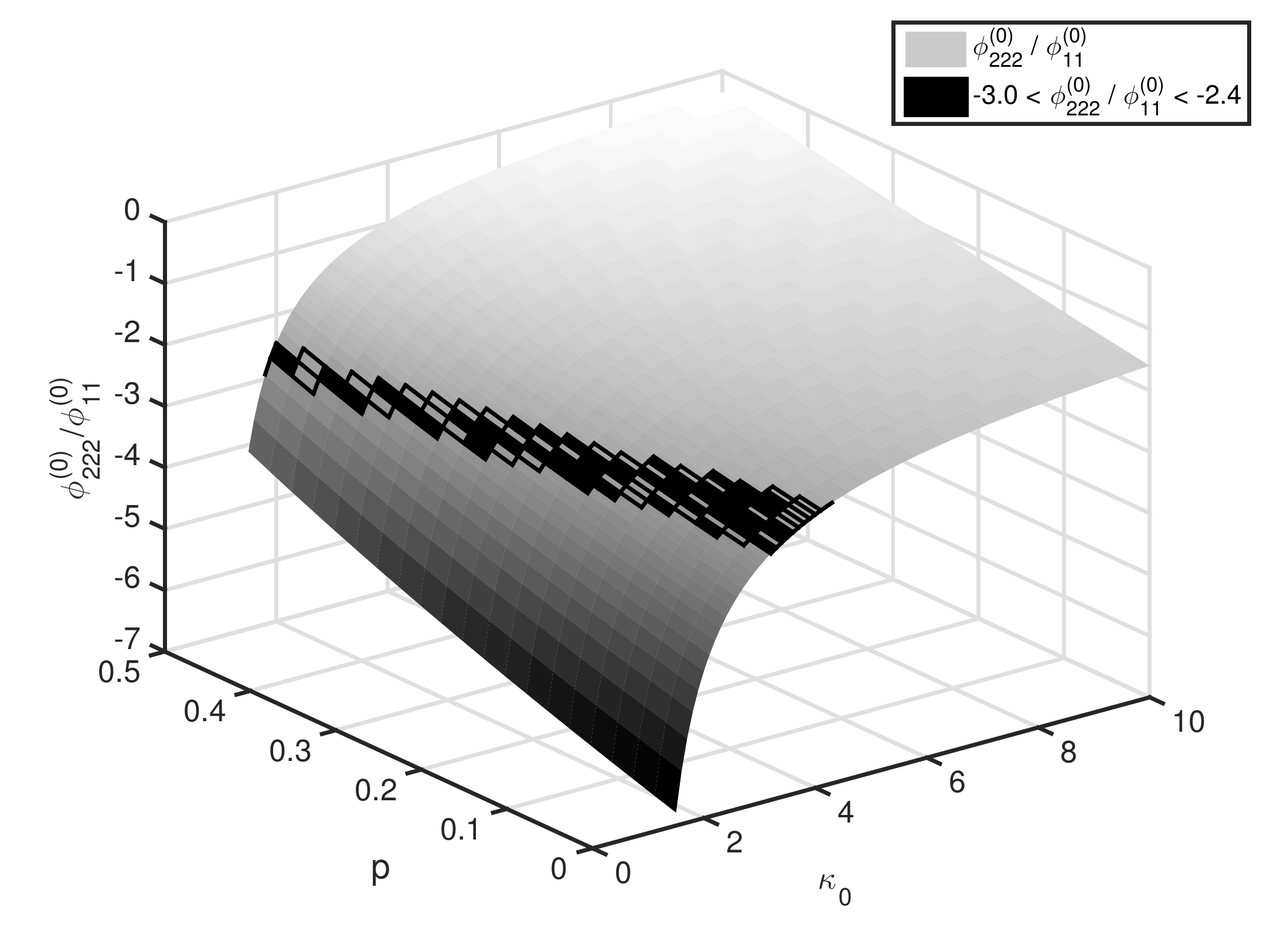}
    \caption{Dependence of $\phi_{222}^{(0)}/\phi_{11}^{(0)}$ on the model parameters $p$ and $\kappa_0$ for the non-singular axisymmetric models, shown in a three dimensional feature space for $p \in \left[0, 0.5 \right]$ and $\kappa_0 \in \left]1, 10 \right]$.}
\label{fig:axisymmetric_degeneracy}
\end{figure}
\noindent
Moreover, to show how model parameters can be obtained in the case of an underdetermined system of equations, we consider the subclass of non-singular axisymmetric models given by
\begin{equation}
  \kappa(x,p) = \kappa_0\dfrac{1 + px^2}{\left( 1 + x^2 \right)^{2-p}}
  \quad\mbox{with}\quad  0 \le p \le \dfrac12\;,
\end{equation}
as defined in \cite{bib:SEF}. For $p=0$, the distribution yields the Plummer model, for $p=1/2$, we obtain a non-singular isothermal sphere. In these two cases, $\kappa_0$ is given by
\begin{equation}
  \kappa_0(x,0) = \dfrac{8GM}{(r_cc)^2}\dfrac{D_\mathrm{ds}}{D_\mathrm{d}D_\mathrm{s}}
  \;,\quad
  \kappa_0 (x,1/2) = \dfrac{4\pi\sigma^2}{r_cc^2}\dfrac{D_\mathrm{ds}}{D_\mathrm{s}}\;,
\end{equation}
where $G$ denotes the gravitational constant, $M$ the total lensing mass, $r_c$ the finite core radius of the lens, and $\sigma^2$ the (measurable) velocity dispersion along the line of sight. The other quantities remain the same as defined in Sect.~\ref{sec:taylor_folds}.

As $x^{(0)}$ is only determined numerically for given values of $p$ and $\kappa_0$, we obtain the ratio of derivatives dependent on $p$ and $\kappa_0$ as shown in Fig.~\ref{fig:axisymmetric_degeneracy} in the parameter range of $p \in \left[0, 1/2 \right]$ and $\kappa_0 \in (1, 10]$. Given measured values for $\phi_{222}^{(0)}/\phi_{11}^{(0)}$, the viable $(p,\kappa_0)$-sets can be read off the graph, as indicated by the black area for the example range of $-3.0 < \phi_{222}^{(0)}/\phi_{11}^{(0)} < -2.4$.

\subsection{Elliptical lens models}
\label{sec:elliptical_lens_models}

Elliptical lens models can be further divided into two classes, elliptical mass distributions and elliptical lensing potentials, as compared in \cite{bib:Kassiola}. For large ellipticities, the latter generate dumb-bell shaped, unrealistic mass distributions, while for small ellipticities an equivalence relation to elliptical mass distributions can be found (see Sect.~5 of \citealt{bib:Kassiola} for details), such that elliptical potentials yield similar observables as elliptical mass distributions. To simplify calculations further, an axi-symmetric primary potential with external shear can also be considered equivalent in many cases of small ellipticities, as stated in \cite{bib:Kovner}.

For the general case of arbitrary ellipticity, we now calculate the model parameters of a singular isothermal ellipse (SIE) as a representative example model of elliptical mass distributions, which we shall test for its suitability to describe the gravitational lensing configurations shown in Sect.~\ref{sec:examples}. The deflection potential of an SIE in polar coordinates is given by
\begin{equation}
  \psi(r,\varphi) = a\sqrt{\tfrac{f}{1-f^2}}r\left(
    |\sin\varphi|\mathrm{acos}(\Delta) + |\cos\varphi|\mathrm{acosh}(\Delta/f)
  \right)
\label{eq:sie_potential}
\end{equation}
with
\begin{equation}
  a = 4 \pi\dfrac{D_{\mathrm{ds}}}{D_\mathrm{s}}\dfrac{\sigma^2}{c^2}\;,\quad
  r = \sqrt{x_1^2 + x_2^2}
\end{equation}
and
\begin{equation}
  \Delta = \sqrt{\cos^2\varphi + f^2 \sin^2\varphi}\;,
\end{equation} 
where $f$ denotes the axis ratio of the semi-minor to the semi-major axis in addition to the quantities already introduced.

Inserting $\psi$ into the lensing potential $\phi(x) = 1/2(x-y)^2 - \psi(x)$ and calculating the derivatives of this lens model as required by Eqs.~\ref{eq:ratio1_cusp} and \ref{eq:ratio2_cusp}, we can use these equations to solve for $a$ and $f$ to obtain
\begin{align}
  a = \dfrac{1}{\sqrt{-r_{122}^{(0)}}\left( r_{2222}^{(0)} - 2\left(r_{122}^{(0)}\right)^2\right)^{1/4}}
  \;,\quad
  f = \dfrac{- r_{122}^{(0)}}{\sqrt{r_{2222}^{(0)} - 2\left(r_{122}^{(0)}\right)^2}} \label{eq:SIE_maj}
\end{align}
for images in the vicinity of a cusp singularity on the semi-major axis of the lens and
\begin{align}
  a = \dfrac{1}{\sqrt{-r_{122}^{(0)}}\left( r_{2222}^{(0)} - 2\left(r_{122}^{(0)}\right)^2\right)^{1/4}}
  \;,\quad
  f = \sqrt{\dfrac{r_{2222}^{(0)} -2\left(r_{122}^{(0)}\right)^2}{\left(r_{122}^{(0)}\right)^2}}
\label{eq:SIE_min}
\end{align}
for images in the vicinity of a cusp singularity on the semi-minor axis of the lens, with $r_{122}^{(0)}$ given by the right-hand side of Eq.~\ref{eq:ratio1_cusp} and $r_{2222}^{(0)}$ by the right-hand side of Eq.~\ref{eq:ratio2_cusp} containing the measured quantities.

\begin{table*}[t]
 \caption{Measured quantities for B1422+231 as summarised in \cite{bib:JVAS}. Ellipticities are taken from \cite{bib:Bradac} and magnification ratios have been calculated from the MERLIN data at 5 GHz radio frequency.}
\label{tab:B1422}
\begin{center}
\begin{tabular}{lrrllllll}
\hline
\noalign{\smallskip}
  Image & $x_1 \left[\text{mas}\right]$ & $x_2 \left[\text{mas}\right]$ & $\Delta x_{1,2} \left[\text{mas}\right]$ & $\mu_{Ai}$ & $\Delta \mu_{Ai}$ & $|\epsilon|$ & $\Delta |\epsilon|$ & Time delay $\left[\text{d}\right]$ \\
\noalign{\smallskip}
\hline
\noalign{\smallskip}
  A & $0.00$       & $0.00$      & $0.05$ & $1.00$ & $0.020$ & $0.80$ & $0.07$ & $t_d^{AB} = 1.5 \pm 1.4$ \smallskip\\
  B & $-389.25$ & $319.98$   & $0.05$ & $0.98$ & $0.020$ & $0.70$ & $0.07$ & $t_d^{BC} = 7.6 \pm 2.5$ \smallskip \\
  C & $333.88$  & $-747.71$  & $0.05$ & $0.52$ & $0.020$ & $0.55$ & $0.09$ & $t_d^{AC} = 8.2 \pm 2.0$ \smallskip \\
  D & $-950.65$ & $-802.15$ & $0.05$ & $0.02$ & $0.005$ & $0.20$ & $0.10$ \\
\noalign{\smallskip}
\hline
\end{tabular}
\end{center}
\end{table*}

\subsection{External shear}
\label{sec:external_shear}

External shear is included into the analysis by adding the term
\begin{equation}
  \phi_\Gamma(x) = \dfrac12\Gamma_1 \left( x_1^2 - x_2^2 \right) + \Gamma_2x_1x_2
\end{equation}
to the lensing potential $\phi(x)$ of the primary gravitational lens, where $\Gamma_i$, $i=1,2$ are real constants. They parametrise the external shear whose orientation $\theta$ and magnitude $\Gamma$ are given by
\begin{equation}
  \theta = \dfrac12 \tan^{-1}\left( \dfrac{\Gamma_2}{\Gamma_1}\right)
  \;,\quad
  \Gamma = \sqrt{\Gamma_1^2 + \Gamma_2^2}\;.
\end{equation}
Since $\phi_\Gamma$ is quadratic in the coordinates, second-order derivatives of the lensing potential change to
\begin{align}
  \phi_{11}(x) &\to \phi_{11}(x) + \Gamma_1\;, \\
  \phi_{22}(x) &\to \phi_{22}(x) - \Gamma_1\;, \\
  \phi_{12}(x) &\to \phi_{12}(x) + \Gamma_2\;,
\end{align}
and all higher-order derivatives remain unchanged. Thus, information from measured time delays is also not affected. This implies that external shear only affects the denominator of the ratios of derivatives in Eqs.~\ref{eq:ratio_fold}, \ref{eq:ratio1_cusp}, and \ref{eq:ratio2_cusp}. For a fixed, measured right-hand side, the convergence of the primary model with external shear changes compared to the convergence of a model without external shear $\kappa(x^{(0)})$ according to
\begin{equation}
\kappa \to \kappa + \dfrac{\Gamma_1}{2}\;,
\end{equation}
now to be taken at the new position of the critical curve after introducing the external shear.

As adding a constant external shear is a global property of the lens mapping, a consistency check can be established by comparing the values of $\Gamma_i$, $=1,2$ determined by several sets of images. For this, the shear values obtained at different singular points have to be aligned by rotation into one global coordinate system.

\section{Examples}
\label{sec:examples}

\begin{figure}
  \centerline{\includegraphics[width=0.7\hsize]{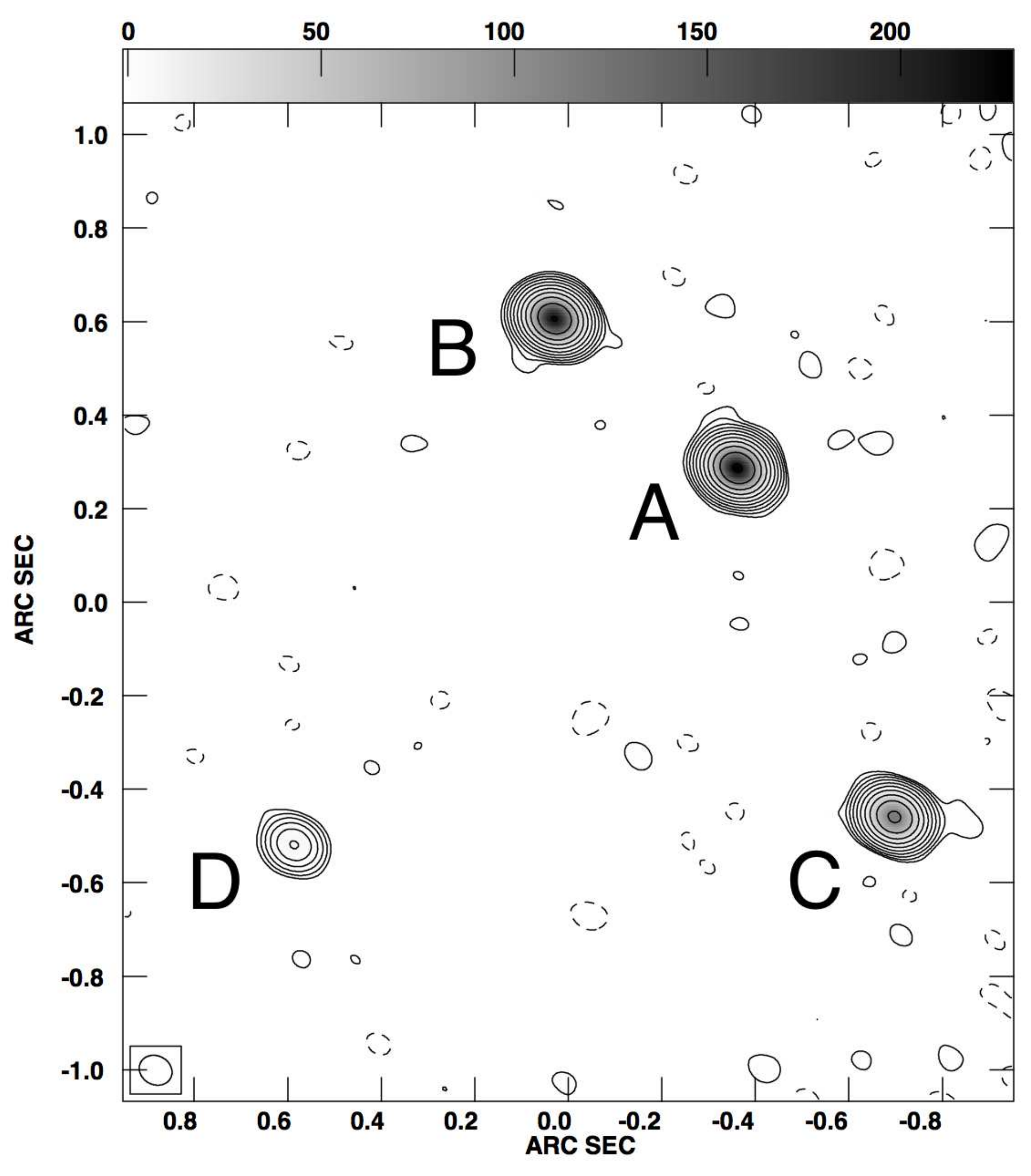}}
  \caption{MERLIN map of B14122+231 at 5 GHz radio frequency, shown here to define our labelling of the four gravitationally lensed images. Images $A$, $B$, and $C$ are close to a cusp singularity, $A$ being closest to the singular point. Image $D$ is on the opposite side and thus not included in our data analysis.}
\label{fig:B1422}
\end{figure}

\subsection{Galaxy lensing -- JVAS B1422+231}
\label{sec:examples_B1422}

JVAS B1422+231, as first described in \cite{bib:B1422}, is a quadruple-image gravitational lens at $z=0.34$ showing three images of a source at $z=3.62$ lying close together, as shown in Fig.~\ref{fig:B1422}. The measured data for this system is summarised in Table~\ref{tab:B1422}. They suggest that the images $A$, $B$ and $C$ originate from a source near a cusp singularity in the caustic. Following our earlier notation, we label the images as shown in Fig.~\ref{fig:B1422}\footnote{Note that common labelling interchanges $A$ and $B$ in Fig.~\ref{fig:B1422}.}.

Since the time delays in Table~\ref{tab:B1422} imply that image $A$ follows both $B$ and $C$, we conclude that $A$ must have negative parity, while $B$ and $C$ must have positive parity (see also \citealt{bib:Congdon}). Applying Eqs.~\ref{eq:time_delay_cusp}, \ref{eq:ratio1_cusp}, and \ref{eq:ratio2_cusp} to the data in Table~\ref{tab:B1422}, we obtain the model-independent information after all observed image positions have been converted to radians
\begin{align}
  -1.622 \le 10^{-5}\,&\dfrac{\phi_{122}^{(0)} }{\phi_{11}^{(0)}}\, \left( \text{rad} \right)^{-1}\le -1.498 \;, \\
  0.12 \le 10^{-12}\,&\dfrac{\phi_{2222}^{(0)} }{\phi_{11}^{(0)}}\, \left( \text{rad} \right)^{-2} \le 1.15 \;, \\
  0.22 \le 10^{-11}\,&\phi_{2222}^{(0)}\, \left( \text{rad} \right)^{-4} \le 1.91 \;, \\
  -3.18 \le 10^{-4}\,&\phi_{122}^{(0)} \, \left( \text{rad} \right)^{-3} \le -2.43 \;, \\
  0.17 \le &\phi_{11}^{(0)} \left( \text{rad}\, \right)^{-2} \le 0.20 \;.
\end{align}
Using Eq.~\ref{eq:SIE_maj} on these ratios, we infer the model parameters of an SIE
\begin{equation}
 2.42 \le 10^6\,a \le 5.01 \;,\quad
 0.14 \le f \le 0.64
\label{eq:SIE_model_parameters_B1422}
\end{equation}
which, solving $a$ for $\sigma$, yields a velocity dispersion of
\begin{equation}
  146.38 \le \sigma\,\left(\mathrm{km\,s^{-1}}\right)^{-1} \le 210.32 \;.
\end{equation}

These parameter values agree well with those found by \cite{bib:Bradac} and \cite{bib:Kormann_B1422}: \cite{bib:Kormann_B1422} determine the velocity dispersion of B1422+231 to be around 200~km/s for axis ratios between $0.35$ and $0.60$, while \cite{bib:Bradac} get an axis ratio of 0.68 with an SIE including external shear and velocity dispersions of 190~km/s. Although being consistent with each other and our results, both methods yield $\chi^2$ values per degree of freedom much larger than unity, rejecting the hypothesis that the resulting model parameters are (locally) optimal.

To assess the quality of our Taylor approximation in this case, we can determine the source position as described in Sect.~\ref{sec:taylor_cusps} to obtain
\begin{equation}
  17.35\,\mathrm{mas} \le y_1 \le 3.74'' \;,\quad
  -1.65'' \le y_2 \le 5.75\,\mbox{mas} \;.
\label{eq:source_positions_B1422}
\end{equation}
Further taking into account that the Einstein radius of the lens is of the order of $1''$, as estimated by the distance between the images $A$ and $D$, a distance of the source to the singular point of the order of 10 mas implies that the Taylor-expanded ratios of derivatives should deviate only by a few percent from their true value, as argued in Sect.~\ref{sec:taylor_error}.

\begin{figure}
  \includegraphics[width=\hsize]{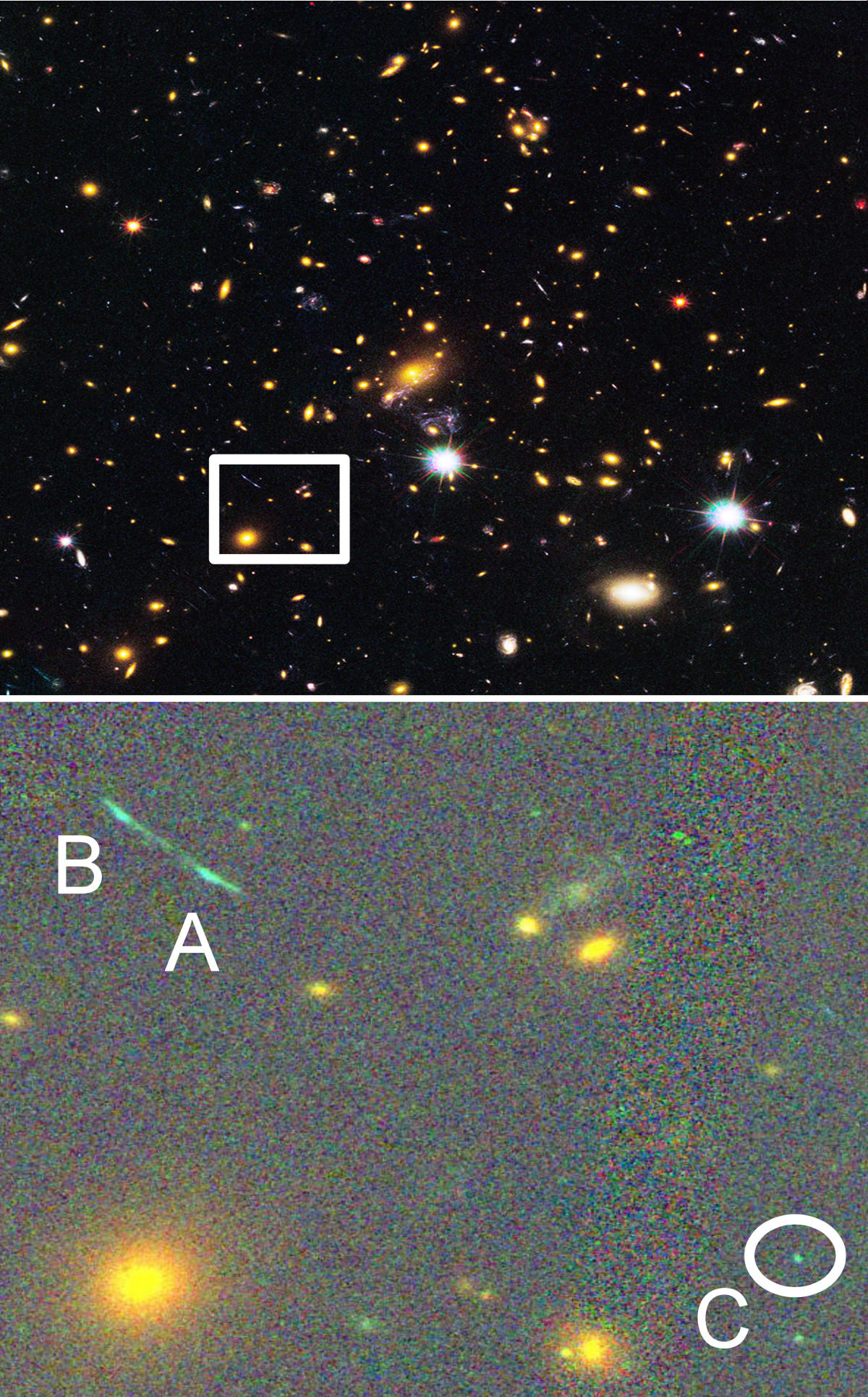}
  \caption{Multi-wavelength image of the galaxy cluster MACS J1149.5+223 taken by the Hubble space telescope (top). The white box marks the position of the three gravitationally lensed images. $A$, $B$, and $C$ (bottom) used for the mass reconstruction within the galaxy cluster. \textit{Image credits: NASA, ESA, and M. Postman (STScI), and the CLASH collaboration}.}
\label{fig:J1149}
\end{figure}

\subsection{Galaxy cluster lensing -- MACS J1149.5+2223}
\label{sec:examples_J1149}

The galaxy cluster MACS J1149.5+223, where a multiply imaged supernova was recently detected \citep{bib:Kelly2} is an X-ray bright, strongly lensing cluster at redshift $z=0.544$, as described in \cite{bib:Ebeling}. For the three images of a source at $z=1.89$ in the right part of Fig.~\ref{fig:J1149}, the CLASH collaboration has determined the distances between the images
\begin{equation}
  \delta_{AB} = 2.42'' \;,\quad \delta_{AC} = 16.25'' \;,\quad \delta_{BC} = 18.66''
\label{eq:J1149_distances}
\end{equation}
with the image ellipticities and their rms-errors obtained from SExtractor
\begin{equation}
  \epsilon_A = 0.685 \pm 0.147 \;,\quad \epsilon_B = 0.686 \pm 0.147 \;,\quad \epsilon_C = 0.128 \pm 1.116\;.
\label{eq:J1149_ellipticities}
\end{equation}
Using the distances of Eq.~\ref{eq:J1149_distances} in radians and the image ellipticities of Eq.~\ref{eq:J1149_ellipticities}, we obtain
\begin{align}
  -2.17 \le 10^{-3}\,&\dfrac{\phi_{122}^{(0)}}{\phi_{11}^{(0)}}\, \left( \text{rad} \right)^{-1} \le -2.16 \;, \\
  0.31 \le 10^{-9}\,&\dfrac{\phi_{2222}^{(0)}}{\phi_{11}^{(0)}}\,  \left( \text{rad} \right)^{-2} \le 1.31 \;.
\end{align}
Lacking time-delay information, we can neither determine the parity of the images nor gain further information as to whether the images are in the vicinity of the cusp on the semi-minor or semi-major axis of the lens. From the observed values, the model parameters for a singularity on the semi-major axis of an SIE are
\begin{eqnarray}
  1.13 \le 10^4\,a \le 1.63 \;,\quad
  0.0599 \le f \le 0.1245 \;,
\label{eq:SIE_major}
\end{eqnarray}
and if the images are in the vicinity of a cusp singularity at the semi-minor axis of an SIE,
\begin{eqnarray}
  1.13 \le 10^4\,a \le 1.63 \;,\quad
  0.0037 \le f \le 0.0077 \;.
\label{eq:SIE_minor}
\end{eqnarray}
From these parameters the velocity dispersion for both cases, Eq.~\ref{eq:SIE_major} and \ref{eq:SIE_minor} is derived to be
\begin{equation}
 1164 \le \sigma\,\left(\mathrm{km\,s^{-1}}\right)^{-1} \le 1397 \;,
\end{equation}
which agrees well with the measured values falling between $500$ and $1270\,\mathrm{km\,s^{-1}}$ from \cite{bib:Smith}.

Due to the different definitions and the addition of further visible mass content to the dark matter halo, there is a large, but consistent range of mass estimates obtained for MACS J1149 of the order $10^{15}M_\odot$ (\cite{bib:Limousin}, \cite{bib:Smith}, \cite{bib:Umetsu}, \cite{bib:Zitrin}), which agrees with the estimated mass of an SIE given the observables for MACS J1149
\begin{align}
  M_{200} &= \dfrac{\pi\sigma^2}{200G}\, D_\mathrm{d} \\
  &\in \left[ 6.56, 9.46 \right]\,10^{15}\,M_{\odot} \;,
\label{eq:mass_SIE}
\end{align}
where $M_{200}$ is the dark halo mass at $r_{200}$, the radius enclosing a mean overdensity of 200 times the critical density of the universe. 

\section{Summary and discussion}
\label{sec:summary}

We have studied here which model-independent characteristics of strong gravitational lenses can be extracted directly from observational data. These observational data include the distances, ellipticities, magnification ratios, and possibly time delays of multiply gravitationally-lensed images of sources close to fold and cusp singularities. Taylor-expanding the lensing potential around these singular points and choosing the coordinate system suitably, we set up a system of non-linear, approximate lensing equations. We solved these equations for the derivatives of the lensing potential at the cusps and folds. As the system is underdetermined even in the leading-order approximation, we could not solve for the derivatives directly, but rather obtained ratios of derivatives. These are connected to physically more intuitive quantities like ratios of flexion and convergence. Time-delay information was used to determine the parities of the images. With given magnification ratios, the source position can be reconstructed, which allows estimating the accuracy of the Taylor expansion of the lensing potential. Furthermore, assuming a specific lens model, we showed that the derivatives of this lens model can be used to determine lens-model parameters. The application of our method to the galaxy-lensing configuration of JVAS B1422+231 and the galaxy-cluster lensing configuration of MACS J1149.5+2223 demonstrated that the model-independent information is capable of reproducing parameter values for an SIE that agree well with measured values and those obtained by $\chi^2$-parameter-estimation.

\begin{acknowledgements}
We wish to thank Mauricio Carrasco, Dan Coe, Matteo Maturi, Sven Meyer, Eberhard Schmitt, Gregor Seidel, Keiichi Umetsu, Gerd Wagner and Leonard Wirsching for helpful discussions. We gratefully acknowledge the support by the Deutsche Forschungsgemeinschaft (DFG) WA3547/1-1.\end{acknowledgements}

\bibliographystyle{aa}
\bibliography{aa}

\appendix

\section{Derivations}

\subsection{Folds}
\label{app:folds}

The Taylor expansions of the derivatives at the centre of light image points $i = A, B$ are given by
\begin{align}
  \phi_{11}^{(i)} &\approx \phi_{11}^{(0)} \phantom{x_{i11}} + \phi_{112}^{(0)}x_{i2} \;, \\
  \phi_{12}^{(i)} &\approx \phi_{112}^{(0)}x_{i1} + \phi_{122}^{(0)}x_{i2} \;, \\
  \phi_{22}^{(i)} &\approx \phi_{122}^{(0)}x_{i1} + \phi_{222}^{(0)}x_{i2}
\end{align}
from which follows 
\begin{align}
  \phi_{11}^{(A)} - \phi_{11}^{(B)} &= \phi_{112}^{(0)}\delta_{AB2} \;, \\
  \phi_{12}^{(A)} - \phi_{12}^{(B)} &= \phi_{122}^{(0)}\delta_{AB2} \;, \\
  \phi_{22}^{(A)} - \phi_{22}^{(B)} &= \phi_{222}^{(0)}\delta_{AB2}
\end{align}
from which can be deduced
\begin{equation}
  r_i \approx \dfrac{\phi_{22}^{(i)}}{\phi_{11}^{(i)}} =
  \dfrac{\phi_{22}^{(i)}}{\phi_{11}^{(0)}} + \mathcal{O}(\delta_{ij}^2)
  \quad  i, j = A, B \quad i \ne j \;.
\end{equation}
Inserting the Taylor expansions into the lensing equations, Eqs.~\ref{eq:fold_le1} and \ref{eq:fold_le2}, we obtain
\begin{align}
  2\phi_{12}^{(A)} &= - 2\phi_{12}^{(B)} = \phi_{122}^{(0)}\delta_{AB2} \;, \label{eq:fold_extended_le1}\\
  2\phi_{22}^{(A)} &= - 2\phi_{22}^{(B)} = \phi_{222}^{(0)}\delta_{AB2} \;. \label{eq:fold_extended_le2}
\end{align}
From Eq.~\ref{eq:fold_extended_le1} we cannot retrieve any information about the ratio of the derivatives, as the equation is also solved by setting $\phi_{12}^{(A)} = \phi_{122}^{(0)} = 0$.
Using $\phi_{22}^{(A)} \approx r_A\phi_{11}^{(A)} \approx r_A \,\phi_{11}^{(0)}$ in Eq.~\ref{eq:fold_extended_le2}, we arrive at
\begin{equation}
  \dfrac{\phi_{222}^{(0)}}{\phi_{11}^{(0)}} = \dfrac{2r_A}{\delta_{AB2}} \;.
\end{equation}

\subsection{Cusps}
\label{app:cusps}

Subtracting the first lensing equation, Eq.~\ref{eq:cusp_le1} for $B$ and $C$ from $A$, respectively, Eq.~\ref{eq:ratio1_cusp} can be immediately obtained. Subsequently, the second lensing equations, Eq.~\ref{eq:cusp_le2}, for the three images are analogously subtracted and the two resulting equations linearised. The Taylor expansions of the second order derivatives of image $i, j = A, B, C$ with $i \ne j$ are
\begin{align}
  \phi_{11}^{(i)} &\approx \phi_{11}^{(0)} \;, \\
  \phi_{12}^{(i)} &\approx \phi_{122}^{(0)}x_{i2} \;, \\
  \phi_{22}^{(i)} &\approx \phi_{122}^{(0)}x_{i1} + \dfrac12 \phi_{2222}^{(0)}x_{i2}^2
\end{align}
from which follows
\begin{align}
  \phi_{12}^{(i)} - \phi_{12}^{(j)} &= \phi_{122}^{(0)}\delta_{ij2} \;, \\
  \phi_{22}^{(i)} - \phi_{22}^{(j)} &= \phi_{122}^{(0)}\delta_{ij1} + \phi_{2222}^{(0)}\delta_{ij2}x_{i2} - \dfrac12 \phi_{2222}^{(0)}(\delta_{ij2})^2 \;, \\
  \phi_{22}^{(i)} - \phi_{22}^{(j)} &\approx \phi_{122}^{(0)}\delta_{ij1} - \dfrac12 \phi_{2222}^{(0)}(\delta_{ij2})^2 \;,
\end{align}
where we used $\phi_{222}^{(i)} \approx  \phi_{2222}^{(0)}x_{i2} \approx \phi_{222}^{(0)} = 0$ in the last step.
Applying these relations to the two resulting, linearised lensing equations yields
\begin{equation}
\dfrac{\phi_{22}^{(A)}}{\phi_{12}^{(A)}}  = - \dfrac{\delta_{AB1}}{\delta_{AB2}} = - \dfrac{\delta_{AC1}}{\delta_{AC2}} \;.
\label{eq:alpha}
\end{equation}
In order to be able to set up this equation, we require that $x_{A2} \ne 0$. This is a reasonable requirement, if the images are not supposed to lie at the singular point. Assuming that $\delta_{AB}$ has an angle $\alpha$ with the $x_1$-axis of the coordinate system, all coordinate distances $\delta_{ijk}$ can be expressed in terms of $\delta_{ij}$, the observed angles $\alpha_i$, $i=A,B,C$, and $\alpha$:
\begin{align}
  \delta_{AB1} &= -\delta_{AB}\cos(\alpha) \;, \\
  \delta_{AB2} &= -\delta_{AB}\sin(\alpha) \;, \\
  \delta_{AC1} &= -\delta_{AC}\cos(\alpha_A - \alpha) \;, \\
  \delta_{AC2} &= \phantom{-}\delta_{AC}\sin(\alpha_A - \alpha) \;,
\end{align}
so that we can solve Eq.~\ref{eq:alpha} for $\alpha$
\begin{equation}
  \alpha = \alpha_A / 2 \quad \text{with} \quad \alpha_A = \pi - \text{acos}\left( \dfrac{\delta_{BC}^2 - \delta_{AB}^2 - \delta_{AC}^2}{2\delta_{AB}\delta_{AC}}\right) \;.
\end{equation}
Subtracting the ratios of two images, $i=A, B, C$
\begin{equation}
  r_i - r_j = \dfrac{\phi_{22}^{(i)}}{\phi_{11}^{(0)}} -\dfrac{\phi_{22}^{(j)}}{\phi_{11}^{(0)}} = \dfrac{\phi_{122}^{(0)}}{\phi_{11}^{(0)}}\delta_{ij1} - \dfrac12\dfrac{\phi_{2222}^{(0)}}{\phi_{11}^{(0)}}(\delta_{ij2})^2
\end{equation}
and inserting Eq.~\ref{eq:ratio1_cusp}, we obtain Eq.~\ref{eq:ratio2_cusp}.

Analogously to the fold case derived in \cite{bib:SEF}, we can derive the time delay to leading order for the cusp between two of the three images $i$ and $j$, with $i, j = A, B, C$, $i \ne j$
\begin{align}
  ct_\mathrm{d}^{(ij)} &= \dfrac{D_\mathrm{d}D_\mathrm{s}}{D_{\mathrm{{ds}}}}
  (1+z_\mathrm{d})(\phi^{(i)} - \phi^{(j)}) \equiv
  \Gamma_\mathrm{d}(\phi^{(i)} - \phi^{(j)}) \\
  &=  - \Gamma_\mathrm{d}\dfrac{(\delta_{ij2})^2}{8\phi_{11}^{(0)}}
  \left(\phi_{11}^{(0)}\phi_{2222}^{(0)} +
  2\left(\phi_{122}^{(0)}\right)^2 \right)(\delta_{ij2})^2  \\
  &+ \Gamma_\mathrm{d}\dfrac{(\delta_{ij2})^2}{2}\phi_{122}^{(0)}\delta_{ij1} \nonumber \\
  &\approx - \Gamma_\mathrm{d}\dfrac{(\delta_{ij2})^4}{8}\phi_{2222}^{(0)}\;.
\label{eq:cusp_time_delay}
\end{align}

\end{document}